\documentclass[apj]{emulateapj}

\usepackage{times}
\usepackage{natbib}
\usepackage[backref,breaklinks,colorlinks,citecolor=cyan]{hyperref} 
\usepackage{booktabs}
\usepackage{graphicx}
\usepackage{bm}
\usepackage{multirow}
\usepackage{enumitem}
\usepackage{amsmath}
\usepackage{float}
\usepackage[caption=false]{subfig}

\shortauthors{Silverman et al.}

\begin{document}


\title{Where do quasar hosts lie with respect to the size - mass relation of galaxies?}


\author{John D. Silverman\altaffilmark{1,2}, Tommaso Treu\altaffilmark{3}, Xuheng Ding\altaffilmark{3, 4}, Knud Jahnke\altaffilmark{5}, Vardha N. Bennert\altaffilmark{6}, Simon Birrer\altaffilmark{3}, Malte Schramm\altaffilmark{7}, Andreas Schulze\altaffilmark{7}, Jeyhan S. Kartaltepe\altaffilmark{8}, David B. Sanders\altaffilmark{9}, Renyue Cen\altaffilmark{10}}
\email{silverman@ipmu.jp}

\altaffiltext{1}{Kavli Institute for the Physics and Mathematics of the Universe, The University of Tokyo, Kashiwa, Japan 277-8583 (Kavli IPMU, WPI)}
\altaffiltext{2}{Department of Astronomy, School of Science, The University of Tokyo, 7-3-1 Hongo, Bunkyo, Tokyo 113-0033, Japan}
\altaffiltext{3}{Department of Physics and Astronomy, University of California, Los Angeles, CA, 90095-1547, USA}
\altaffiltext{4}{School of Physics and Technology, Wuhan University, Wuhan 430072, China}
\altaffiltext{5}{Max-Planck-Institut f\"ur Astronomie, K\"onigstuhl 17, D-69117 Heidelberg, Germany}
\altaffiltext{6}{Physics Department, California Polytechnic State University, San Luis Obispo CA 93407, USA}
\altaffiltext{7}{National Astronomical Observatory of Japan, Mitaka, Tokyo 181-8588, Japan}
\altaffiltext{8}{School of Physics and Astronomy, Rochester Institute of Technology, 84 Lomb Memorial Drive, Rochester, NY 14623, USA}
\altaffiltext{9}{Institute for Astronomy, University of Hawaii, 2680 Woodlawn Drive, Honolulu, HI, 96822}
\altaffiltext{10}{Department of Astrophysical Sciences, Princeton University, Princeton, NJ 08544-1001, USA}



\begin{abstract}
The evolution of the galaxy size -- stellar mass (M$_{stellar}$) relation has been a puzzle
for over a decade. High redshift galaxies are significantly more compact than galaxies observed todayâ at an equivalent mass, but how much of this apparent growth is driven
by progenitor bias, minor mergers, secular processes, or feedback from
active galactic nuclei (AGN) is unclear. To help disentangle the physical
mechanisms at work by addressing the latter, we study the size - M$_{stellar}$ relation of 32 carefully-selected broad-line AGN hosts at $1.2<z<1.7$ ($7.5<\log~M_{BH}<8.5$; $L_{bol}/L_{Edd} \gtrsim 0.1$). Using HST with multi-band photometry and state-of-the-art modeling techniques, we measure
half-light radii while accounting for uncertainties from subtracting bright
central point sources. We find AGN hosts to have sizes ranging from 1 to 6 kpc at $M_{stellar}\sim0.3-1\times10^{11}$ M$_{\odot}$. Thus, many hosts have intermediate sizes as compared
to equal-mass star-forming and quiescent galaxies. While inconsistent
with the idea that AGN feedback may induce an increase in galaxy sizes, this finding
is consistent with hypotheses in which AGNs preferentially occur in systems
with prior concentrated gas reservoirs, or are involved in secular compaction
processes perhaps responsible for simultaneously building bulges and shutting
down star formation. If driven by minor mergersâ which do not grow central
black holes as fast as they do bulge-like stellar structures, such a process
would explain both the galaxy size -- mass relation observed here and the evolution in
the black holeâ bulge mass relation described in a companion paper.

\end{abstract}



\keywords{}






\section{Introduction}

The growth in size of the observed galaxy population with cosmic time is a key observational quantity to formulate a global picture of galaxy evolution \citep[e.g.,][]{Carollo2013,vanderWel2014,Barro2017,Faisst2017}. Both star-forming and passive galaxies exhibit an increase in their effective radii with declining redshift, along parallel tracks with the passive galaxies being more compact, even out to high redshift \citep{Daddi2005,vanDokkum2008,Bezanson2009}. 

Explaining the change in size for both star-forming and quiescent galaxies with redshift has been a challenge from a galaxy formation standpoint. A variety of physical processes have been invoked but no consensus has emerged yet on a global self consistent picture. One of the ingredients is progenitor bias, where the galaxy population at high-$z$ is not the full set of progenitors of today's galaxies but most likely a denser subset owing to initial conditions and perhaps the denser environment in which the earlier galaxies were formed \citep[e.g.][]{Morishita2017}.

Other explanations involve minor mergers or galactic secular processes. For example, the change in size may be attributed to gas inflow onto galaxies that adds angular momentum in the outskirts of star-forming disks \citep[e.g.,][]{Peng2019}. The inside-out growth \citep[e.g.,][]{Tacchella2016} of galaxies can also contribute to the increase in effective galaxy size through star formation persisting at larger radii. In the case of the passive galaxies, other processes such as minor mergers have been put forward to explain their size evolution \citep[e.g.,][]{Naab2009}, although the timescales seem faster than expected \citep[e.g.,][]{Newman2012}, and the scatter difficult to explain \citep[e.g.,][]{Nipoti2012}. It has been suggested that the change in size of both populations is due to a ``compaction'' phase  \citep{DekelBurkert2014}, although considerable debate remains \citep[e.g.,][]{Abramson2018}.

In addition to considering such morphological changes, galaxy evolution models need to consider the evolutionary pathway for galaxies to transition from active to passive states in their star-forming activity, and the role played by feedback from supermassive black holes. It has been suggested that quasar feedback can remove substantial amounts of gas from the inner regions of their host galaxy thus causing an increase in the size of their stellar distribution \citep{Fan2008,Fan2010}. Alternatively, feedback from Active Galactic Nuclei (AGN) can induce star formation within the AGN-driven outflow thus stars are formed on larger scales \citep{Ishibashi2013,Ishibashi2014}. To date, there is little observational evidence for such feedback mechanisms as playing a dominant role in the size growth of galaxies. Even so, there is much interest in determining the galaxy state (i.e., star-forming vs. quiescent; disk vs. bulge) for which SMBHs are primarily gaining their mass since there is a tight relation between the mass of a SMBH and the stellar mass of its host in the local Universe \citep[e.g.,][]{Haering2004,Bennert2010} that seems to persist out to high redshifts (in total stellar mass for the latter).

While there are some studies of the sizes of AGN hosts at high redshift \citep{Barro2014,Rangel2014,Kocevski2017}, little is known about the size -- M$_{stellar}$ relation for the more luminous AGNs (i.e. broad-line QSOs), owing to the challenges of separating the host galaxy from the bright point source which requires accurate characterization of the point spread function (PSF).  However, if the technical challenges can be overcome, a comparison between the size -- M$_{stellar}$ relation of AGN host galaxies and that of the typical galaxy population may shed light particularly on the connection between the growth of quiescent galaxies and SMBHs. 

To establish the relation between the mass of SMBHs and the stellar mass of their host galaxies, we have been conducting an imaging survey of 32 broad-line (type 1) AGNs at $1.2 < z < 1.7$ in deep survey fields (i.e., COSMOS, SXDS and CDF-S) using HST/WFC3 \citep[][; D19 hereafter]{Ding2019} in the near-infrared. By virtue of unprecedented data and state-of-the-art techniques, we have detected the hosts in essentially all cases and measured properties of the host galaxies (i.e., luminosity, size, Sersic index and stellar mass). The key result of our study so far is that the total galaxy stellar mass - SMBH mass relation can be consistent with low-$z$ results, once uncertainties and selection effects are taken into account, but the $bulge$ - SMBH mass relation is not.

In this letter, we investigate the galaxy size - mass relation of type 1 AGNs using our sample at $1.2 < z < 1.7$ and compare with published relations for the general population both star-forming and quiescent.  We show that AGN hosts have sizes between those of star forming and quiescent galaxies at the same stellar mass. We argue that this finding does not provide evidence for the scenarios in which AGN activity is responsible for the growth in size of galaxies. A scenario in which AGN hosts are getting more compact due to the growth of the pressure supported component either by gas rich secular processes or minor mergers seems consistent with the data. Using measurements from \citet{Bennert2011}, we find consistent results with type 1 AGNs at low-$z$. Throughout this paper we use a Hubble constant of $H_0 = 70$ km s$^{-1}$ Mpc$^{-1}$ and cosmological density parameters $\Omega_\mathrm{m} = 0.3$ and $\Omega_\Lambda = 0.7$. We assume a Chabrier initial mass function for estimates of stellar mass. 

\section{Method}

We have carried out an HST/WFC3 IR program in Cycle 25 (PI Silverman) to image the host galaxies of 32 type 1 AGN at $1.2<z<1.7$ in deep survey fields. These AGN have black hole masses ($7.5<\log~M_{BH} < 8.5$), located below the knee of the black hole mass function at their respective redshift, determined from the broad H$\alpha$ emission line detected with Subaru's Fiber Multi-Object Spectrograph (FMOS) as reported in \citet{Schulze2018}. The Eddington ratios are mainly above 0.1 (see Figure 1 of D19). The primary aim of the program is to establish the $M_{BH}-M_{stellar}$ relation, including an inference of the bulge component, at high-$z$ and determine whether there is any evolution in the mass scaling relations by comparing to local values including both inactive and active galaxies.

The procedure to measure stellar mass of the host galaxy requires a decomposition of the total infrared emission into the AGN and host galaxy component through a forward-modeling, chi-squared minimization procedure using the tools available in the \textit{Lenstronomy} image analysis package \citep{Birrer2018}. The inputs are the science frames, 2D PSF models and pixel-level error maps. The host galaxies are modeled as a Sersic function parameterized by an index ($n_{Sersic}$) descriptive of the radial dependence of the light profile and the half-light radius (R$_{\rm eff}$; semi-major axis). The AGN component is fit using model PSFs based on a stellar library constructed from the same WFC3 data set. Based on our analysis as fully presented in D19, we detect the host galaxy in all 32 AGNs with widely varying host-to-total flux ratios with the majority between 20-60\%. Errors are derived from the 1$\sigma$ standard deviation of measurements based on the top eight best-fitting PSF models to the data. The 2D model fits and 1D surface brightness distributions are shown in D19 (Figure 2 and Appendix).

The majority of our sample (21/32) has optical HST imaging available from the COSMOS program. This allows us to perform the image decomposition in two HST bands that bracket the 4000 Angstrom break thus allowing an estimate of the rest-frame color to facilitate accurate stellar mass measurements. A 1 ($z<1.44$) and 0.625 ($z>1.44$) Gyr single stellar population model \citep{Bruzual2003} with solar metallicity appears to nicely fit the two-band HST photometry of the host galaxy (see Figure 5 of D19). These ages of the stellar population for AGN hosts are in good agreement with earlier complementary studies \citep{Jahnke2004, Sanchez2004}. We use this SED to apply a mass-to-light conversion ($z<1.44$: M/L = 0.54; $z > 1.44$: M/L = 0.42) to achieve stellar masses for our full sample. The mass-to-light conversion carries the typical uncertainties for determinations based on a single color \citep[e.g.,][]{B+D01}. Since the fit is not unique to this single stellar population model, one should use caution when further interpreting the stellar age. 

\begin{figure}
\epsscale{1.4}
\plotone{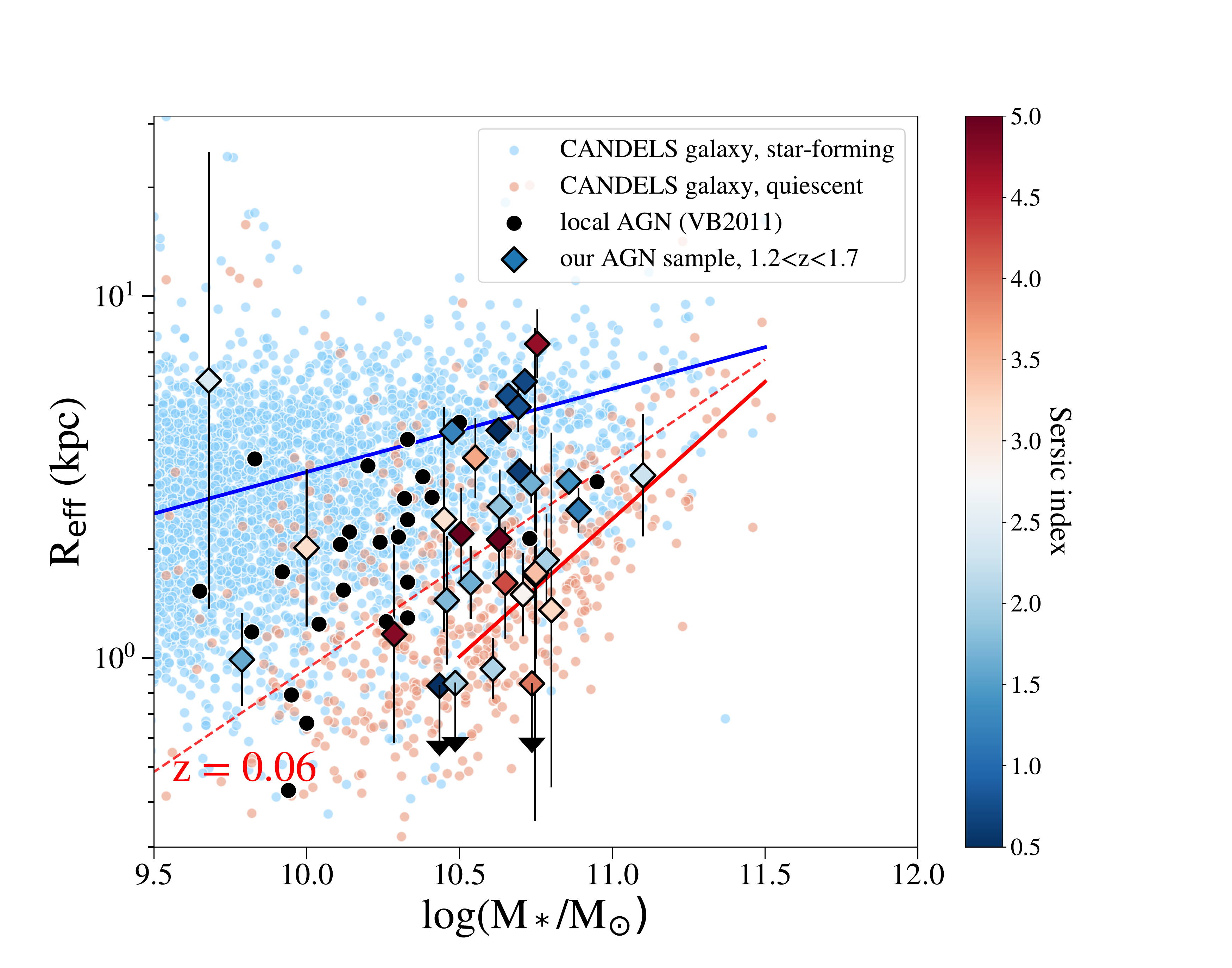}
\caption{Galaxy size - stellar mass relation for the host galaxies of broad-line AGN at $1.2 < z < 1.7$. Our high-$z$ sample is displayed with diamond symbols and a color descriptive of their Sersic index. Arrows indicate those with upper limits for three cases. For comparison, star-forming (blue) and quiescent (red) galaxies from CANDELS are plotted as small circles with a classification based on their rest-frame $U-V$ and $V-J$ colors. The best-fit relations from \citet{vanderWel2014} are shown for the star-forming (blue line) and quiescent (red line) galaxies separately with the latter also indicated at $z\sim0.06$ \citep{Newman2012}. Low-redshift AGNs are marked by the small black circles \citep{Bennert2011}.}
\label{fig:size_mass}
\end{figure}

\section{Results}

The distribution of the parameters descriptive of the properties of type 1 AGN hosts at $1.2 < z < 1.7$ are broad (Fig. 4 of D19). The Sersic index (n$_{Sersic}$) spans the full range of allowed parameter space (0.5 to 6.0) and has a mean of 2.0 while the effective radius (R$_{\rm eff}$) is generally between 1 and 6 kpc with a mean of 2.2 kpc. In general, these values are indicative of a significant disk-like population with rather small sizes as compared to star-forming galaxies at equivalent stellar mass, both local and at these redshifts. While the sizes are in good agreement with lower redshift studies of AGN hosts \citep[][]{Sanchez2004}, the fraction of disks is typically higher.

With a focus here on the galaxy size - M$_{stellar}$ relation, we plot the distribution of $log~\rm{R}_{\rm eff}$ and M$_{stellar}$ in Figure~\ref{fig:size_mass}. For comparison, we also include the individual measurements for the general galaxy population at equivalent redshifts from the CANDELS survey \citep{vanderWel2012,vanderWel2014} along with the best-fit relations for star-forming and quiescent galaxies separately that have been classified through color-color diagrams utilizing the IR photometry (i.e., UVJ diagram). We find that there is a wide spread in size at a given stellar mass for our sample (colored diamonds). While there are a few AGN hosts that fall along the star-forming relation, a fair number are consistent with the quiescent population. The majority of the AGN-host sample falls between the two size - mass relations, thus these galaxies may be undergoing a transition. Based on a Kolmogorov-Smirnov test for galaxies having $10.2\lesssim \log~M_{stellar}<11$, there is a probability of 0.001 that the size distribution of our AGN sample could be drawn from either the star-forming or quiescent galaxies separately.       

It is worth highlighting that our selection of lower mass black holes ($7.5\lesssim \log~M_{BH}<8.5$) in deep survey fields (see Fig. 1 of D19), as opposed to the more massive black holes associated with SDSS quasars, results in a sample having host galaxy masses within a range ($10.5\lesssim \log~M_{stellar}<11$) for which there is a discernible difference between the size - mass relation of star-forming and quiescent galaxies separately. If our sample had higher host stellar masses (e.g., $\log~M_{stellar}\gtrsim11$), the errors on the sizes would make it difficult to carry out these comparisons due to the convergence of the size - mass relations for star-forming and quiescent galaxies. At even lower stellar masses, the quiescent population at high-$z$ within CANDELS is limited in size to carry out such analysis. In any case, our current observations of the SMBH population at these redshifts do not probe black hole masses below $10^{7.5}$ M$_{\odot}$ that would be required.  

We can further investigate whether our results at high redshift are seen in the low redshift Universe using the sample of AGNs at $0.02<z <0.09$ from \citet{Bennert2011}. In Figure~\ref{fig:size_mass}, we find that the low redshift AGNs (small black circles) extend to lower stellar mass ($\sim10^{10}$ M$_{\odot}$) than our sample. Their sizes are systematically elevated from the local size -- M$_{stellar}$ relation for quiescent galaxies and below that of star-forming galaxies. Therefore, we find consistent results at high and low redshift that AGN hosts have sizes in between star-forming (disk-dominated) and quiescent (bulge-dominated) galaxies.

\section{Discussion}

First, we address the ideas that quasars remove gas through feedback that can induce an increase in the stellar size of galaxies \citep{Fan2008,Fan2010} and star formation may occur within AGN outflows thus forming stars on larger scales \citep{Ishibashi2013,Ishibashi2014}. Under both scenarios, one would expect that the hosts of luminous AGNs would start to show evidence for increased sizes relative to an underlying galaxy population without experiencing a luminous AGN phase. Since the Sersic indices of the sample more resemble that of disk galaxies (Fig.~\ref{fig:size_mass}), we could expect their host sizes to be elevated from the mass - size relation of star-forming galaxies. On the contrary, their sizes are considerably smaller. 

Under a scenario where AGN feedback plays a major role in the sizes of galaxies, \citet{Ishibashi2014} argue that the coupling between the black hole and its host should be primarily with the spheroidal component. Thus, the sizes of AGN hosts would follow the size -- mass relation of the quiescent population (or below if feedback effects have not concluded). Counter to the predictions from the model, we find that our AGN sample shows larger sizes relative to the size -- mass relation for quiescent galaxies. This is also the case for the low-redshift AGNs (Fig.~\ref{fig:size_mass}; black circles). If AGN feedback was enabling star formation on larger scales, we would expect AGN hosts to have an extended stellar envelope that would raise their Sersic indices to values comparable to spheroidal galaxies (n$_{Sersic}\sim4$). On the contrary, we find many AGN hosts at high-$z$ having Sersic indices (Figure~\ref{fig:size_mass}) more consistent with disk-like galaxies (n$_{Sersic}\sim2$) and their bulges are under massive for their respective black hole mass (D19). Furthermore, we do not see a direct relation between size and stellar mass for AGN hosts, as expected from the model, likely due to a significant amount of intrinsic dispersion in their observed sizes.

It also is unlikely that the hosts of our AGN sample have had their star formation quenched through feedback effects since we find that the best-fit SED is consistent with a relatively young stellar population (D19). Furthermore, the star formation rates seen in AGN hosts at similar redshifts are consistent with typical star-forming galaxies and not the quiescent population \citep[e.g.,][]{Scholtz2018,Schulze2019}. A direct measure of the star formation rates of our sample is needed to further test this argument. We conclude that there is not any evidence to support claims that a luminous AGN phase has an impact on the sizes of the stellar mass distribution in galaxies at high redshift. It is possible that the full effects of AGN feedback (if present) have not manifested themselves in a change of galaxy size that would occur on longer time scales. These are important issues that can be addressed in the future with observations of the molecular gas, particularly in the central regions. 

A more plausible scenario is that these galaxies are undergoing a structural transition from disk-like to bulge-like stellar distributions. One idea that has been described in the literature is a compaction phase \citep{DekelBurkert2014} that builds the central mass concentration under considerable angular momentum loss. This may involve secular processes (i.e., clump formation and migration) where major mergers are not required to build a bulge \citep{Bournaud2014}. Actually, our sample does not exhibit strong signs of major mergers thus the internal changes occurring are likely due to secular processes such as dynamical instabilities that could be stimulated by gas accretion, minor mergers or even stronger major mergers in the more distant past that are no longer easily discernible. This scenario has been put forward by \citet{Kocevski2017} to explain observational results based on high-z AGNs from deep X-ray survey fields that show a higher fraction of AGNs in compact blue galaxies \citep[][]{Silverman2008,Kocevski2017,Ni2019}. These results are further supported by a recent study of the AGN fraction of compact galaxies \citep{Habouzit2019} using the large cosmological hydrodynamic simulation IllustrisTNG.  

This picture is also consistent with recent ALMA studies of the ISM of galaxies at similar redshifts. It has now been demonstrated that the molecular gas and dust distributions \citep[i.e.,][]{Puglisi2019,Rujopakarn2019} in high-z galaxies are more compact than their stellar distribution even for main-sequence galaxies. Therefore, there is some mechanism(s) that is increasing the central gas density that likely plays a role in forming bulge stars in situ. It may be that our sample illuminates such galaxies since the elevated central gas density is likely to also fuel a SMBH. In fact, many of the galaxies in the sample of \citet{Puglisi2019} have X-ray detected AGNs. Therefore, our type 1 quasars may be signposts for galaxies having elevated central gas densities hence significant growth of not only the SMBH but the bulge \citep{Rujopakarn2018}. However, this structural transition from disk-like to bulge-like stellar distributions could also be fostered by repeated minor mergers not involving gas that can morph the disk into a bulge or simply grow the spheroidal component \cite[e.g.,][]{Croton2006,Nipoti2012}.

\section{Concluding remarks}

At their respective stellar mass, the host galaxies of actively accreting SMBH at $z\sim1.5$ have stellar sizes spanning a range ($\sim$1 - 6 kpc) from the larger star-forming disks to the more compact quiescent galaxies. This result brings into question the role of SMBHs in the structural transformation of galaxies from disks to bulges where a central mass concentration has been tied to the quenching of star formation \citep[e.g.,][]{Tacchella2018}. We discuss scenarios in which the SMBH may be a direct or indirect consequence of an increase in the central mass concentration. We recognize that evidence for AGN hosts to be in a phase of contraction is unsubstantiated. However, it does appear that AGN feedback is unlikely to play a role in the size growth of galaxies with cosmic time. Their sizes would be larger than measured if AGN feedback were effectively removing large amounts of gas or inducing star formation within an outflow thus depositing stars on larger scales. As mentioned above, the time scales for such expansion of their stellar distribution may be longer than probed here.

To more firmly connect the growth between SMBHs and bulges, knowledge of the central gas density will be invaluable for samples such as the one investigated here with stellar mass determinations and information on their stellar bulge component. With a likely mass deficit in their bulges (D19), it is important to identify the physical mechanisms (e.g., minor mergers, large-scale gas accretion, internal disk instabilities) responsible for aligning high-z SMBHs and their host onto the local relation, and their galaxy size - mass distribution. 

\acknowledgments

Based in part on observations made with the NASA/ESA Hubble Space Telescope, obtained at the Space Telescope Science Institute, which is operated by the Association of Universities for Research in Astronomy, Inc., under NASA contract NAS 5-26555. These observations are associated with programs \#15115. Support for this work was provided by NASA through grant number HST-GO-15115 from the Space Telescope Science Institute, which is operated by AURA, Inc., under NASA contract NAS 5-26555. We thank Louis Abramson for useful discussions. XD, SB and TT acknowledge support by the Packard Foundation through a Packard Research fellowship to TT. VNB gratefully acknowledges assistance from a NASA grant associated with HST proposal GO 15215. JS is supported by JSPS KAKENHI Grant Number JP18H01251 and the World Premier International Research Center Initiative (WPI), MEXT, Japan.

\bibliographystyle{apj.bst}
\bibliography{references}

\begin{thebibliography}{}
\expandafter\ifx\csname natexlab\endcsname\relax\def\natexlab#1{#1}\fi

\bibitem[{{Abramson} \& {Morishita}(2018)}]{Abramson2018}
{Abramson}, L.~E., \& {Morishita}, T. 2018, \apj, 858, 40

\bibitem[{{Barro} {et~al.}(2014){Barro}, {Faber}, {P{\'e}rez-Gonz{\'a}lez},
  {Pacifici}, {Trump}, {Koo}, {Wuyts}, {Guo}, {Bell}, {Dekel}, {Porter},
  {Primack}, {Ferguson}, {Ashby}, {Caputi}, {Ceverino}, {Croton}, {Fazio},
  {Giavalisco}, {Hsu}, {Kocevski}, {Koekemoer}, {Kurczynski}, {Kollipara},
  {Lee}, {McIntosh}, {McGrath}, {Moody}, {Somerville}, {Papovich}, {Salvato},
  {Santini}, {Tal}, {van der Wel}, {Williams}, {Willner}, \&
  {Zolotov}}]{Barro2014}
{Barro}, G., {Faber}, S.~M., {P{\'e}rez-Gonz{\'a}lez}, P.~G., {et~al.} 2014,
  \apj, 791, 52

\bibitem[{{Barro} {et~al.}(2017){Barro}, {Faber}, {Koo}, {Dekel}, {Fang},
  {Trump}, {P{\'e}rez-Gonz{\'a}lez}, {Pacifici}, {Primack}, \&
  {Somerville}}]{Barro2017}
{Barro}, G., {Faber}, S.~M., {Koo}, D.~C., {et~al.} 2017, \apj, 840, 47

\bibitem[{{Bell} \& {de Jong}(2001)}]{B+D01}
{Bell}, E.~F., \& {de Jong}, R.~S. 2001, The Astrophysical Journal, 550, 212

\bibitem[{{Bennert} {et~al.}(2011){Bennert}, {Auger}, {Treu}, {Woo}, \&
  {Malkan}}]{Bennert2011}
{Bennert}, V.~N., {Auger}, M.~W., {Treu}, T., {Woo}, J.-H., \& {Malkan}, M.~A.
  2011, \apj, 742, 107

\bibitem[{{Bennert} {et~al.}(2010){Bennert}, {Treu}, {Woo}, {Malkan}, {Le
  Bris}, {Auger}, {Gallagher}, \& {Blandford}}]{Bennert2010}
{Bennert}, V.~N., {Treu}, T., {Woo}, J., {et~al.} 2010, \apj, 708, 1507

\bibitem[{{Bezanson} {et~al.}(2009){Bezanson}, {van Dokkum}, {Tal},
  {Marchesini}, {Kriek}, {Franx}, \& {Coppi}}]{Bezanson2009}
{Bezanson}, R., {van Dokkum}, P.~G., {Tal}, T., {et~al.} 2009, \apj, 697, 1290

\bibitem[{{Birrer} \& {Amara}(2018)}]{Birrer2018}
{Birrer}, S., \& {Amara}, A. 2018, Physics of the Dark Universe, 22, 189

\bibitem[{{Bournaud} {et~al.}(2014){Bournaud}, {Perret}, {Renaud}, {Dekel},
  {Elmegreen}, {Elmegreen}, {Teyssier}, {Amram}, {Daddi}, {Duc}, {Elbaz},
  {Epinat}, {Gabor}, {Juneau}, {Kraljic}, \& {Le Floch'}}]{Bournaud2014}
{Bournaud}, F., {Perret}, V., {Renaud}, F., {et~al.} 2014, \apj, 780, 57

\bibitem[{{Bruzual} \& {Charlot}(2003)}]{Bruzual2003}
{Bruzual}, G., \& {Charlot}, S. 2003, \mnras, 344, 1000

\bibitem[{{Carollo} {et~al.}(2013){Carollo}, {Bschorr}, {Renzini}, {Lilly},
  {Capak}, {Cibinel}, {Ilbert}, {Onodera}, {Scoville}, {Cameron}, {Mobasher},
  {Sanders}, \& {Taniguchi}}]{Carollo2013}
{Carollo}, C.~M., {Bschorr}, T.~J., {Renzini}, A., {et~al.} 2013, \apj, 773,
  112

\bibitem[{{Croton}(2006)}]{Croton2006}
{Croton}, D.~J. 2006, Monthly Notices of the Royal Astronomical Society, 369,
  1808

\bibitem[{{Daddi} {et~al.}(2005){Daddi}, {Renzini}, {Pirzkal}, {Cimatti},
  {Malhotra}, {Stiavelli}, {Xu}, {Pasquali}, {Rhoads}, {Brusa}, {di Serego
  Alighieri}, {Ferguson}, {Koekemoer}, {Moustakas}, {Panagia}, \&
  {Windhorst}}]{Daddi2005}
{Daddi}, E., {Renzini}, A., {Pirzkal}, N., {et~al.} 2005, \apj, 626, 680

\bibitem[{{Dekel} \& {Burkert}(2014)}]{DekelBurkert2014}
{Dekel}, A., \& {Burkert}, A. 2014, \mnras, 438, 1870

\bibitem[{{Ding} {et~al.}(2019){Ding}, {Silverman}, {Treu}, {Schulze},
  {Schramm}, {Birrer}, {Park}, {Jahnke}, {Bennert}, {Kartaltepe}, {Koekemoer},
  {Malkan}, \& {Sanders}}]{Ding2019}
{Ding}, X., {Silverman}, J., {Treu}, T., {et~al.} 2019, arXiv e-prints,
  arXiv:1910.11875

\bibitem[{{Faisst} {et~al.}(2017){Faisst}, {Carollo}, {Capak}, {Tacchella},
  {Renzini}, {Ilbert}, {McCracken}, \& {Scoville}}]{Faisst2017}
{Faisst}, A.~L., {Carollo}, C.~M., {Capak}, P.~L., {et~al.} 2017, \apj, 839, 71

\bibitem[{{Fan} {et~al.}(2010){Fan}, {Lapi}, {Bressan}, {Bernardi}, {De Zotti},
  \& {Danese}}]{Fan2010}
{Fan}, L., {Lapi}, A., {Bressan}, A., {et~al.} 2010, \apj, 718, 1460

\bibitem[{{Fan} {et~al.}(2008){Fan}, {Lapi}, {De Zotti}, \& {Danese}}]{Fan2008}
{Fan}, L., {Lapi}, A., {De Zotti}, G., \& {Danese}, L. 2008, \apjl, 689, L101

\bibitem[{{Habouzit} {et~al.}(2019){Habouzit}, {Genel}, {Somerville},
  {Kocevski}, {Hirschmann}, {Dekel}, {Choi}, {Nelson}, {Pillepich}, {Torrey},
  {Hernquist}, {Vogelsberger}, {Weinberger}, \& {Springel}}]{Habouzit2019}
{Habouzit}, M., {Genel}, S., {Somerville}, R.~S., {et~al.} 2019, \mnras, 484,
  4413

\bibitem[{{H{\"a}ring} \& {Rix}(2004)}]{Haering2004}
{H{\"a}ring}, N., \& {Rix}, H.-W. 2004, \apjl, 604, L89

\bibitem[{{Ishibashi} \& {Fabian}(2014)}]{Ishibashi2014}
{Ishibashi}, W., \& {Fabian}, A.~C. 2014, \mnras, 441, 1474

\bibitem[{{Ishibashi} {et~al.}(2013){Ishibashi}, {Fabian}, \&
  {Canning}}]{Ishibashi2013}
{Ishibashi}, W., {Fabian}, A.~C., \& {Canning}, R.~E.~A. 2013, \mnras, 431,
  2350

\bibitem[{{Jahnke} {et~al.}(2004){Jahnke}, {S{\'a}nchez}, {Wisotzki}, {Barden},
  {Beckwith}, {Bell}, {Borch}, {Caldwell}, {H{\"a}ussler}, {Heymans}, {Jogee},
  {McIntosh}, {Meisenheimer}, {Peng}, {Rix}, {Somerville}, \&
  {Wolf}}]{Jahnke2004}
{Jahnke}, K., {S{\'a}nchez}, S.~F., {Wisotzki}, L., {et~al.} 2004, \apj, 614,
  568

\bibitem[{{Kocevski} {et~al.}(2017){Kocevski}, {Barro}, {Faber}, {Dekel},
  {Somerville}, {Young}, {Williams}, {McIntosh}, {Georgakakis}, \&
  {Hasinger}}]{Kocevski2017}
{Kocevski}, D.~D., {Barro}, G., {Faber}, S.~M., {et~al.} 2017, \apj, 846, 112

\bibitem[{{Morishita} {et~al.}(2017){Morishita}, {Abramson}, {Treu}, {Vulcani},
  {Schmidt}, {Dressler}, {Poggianti}, {Malkan}, {Wang}, {Huang}, {Trenti},
  {Brada{\v{c}}}, \& {Hoag}}]{Morishita2017}
{Morishita}, T., {Abramson}, L.~E., {Treu}, T., {et~al.} 2017, \apj, 835, 254

\bibitem[{{Naab} {et~al.}(2009){Naab}, {Johansson}, \& {Ostriker}}]{Naab2009}
{Naab}, T., {Johansson}, P.~H., \& {Ostriker}, J.~P. 2009, \apjl, 699, L178

\bibitem[{{Newman} {et~al.}(2012){Newman}, {Ellis}, {Bundy}, \&
  {Treu}}]{Newman2012}
{Newman}, A.~B., {Ellis}, R.~S., {Bundy}, K., \& {Treu}, T. 2012, \apj, 746,
  162

\bibitem[{{Ni} {et~al.}(2019){Ni}, {Yang}, {Brandt}, {Alexander}, {Chen},
  {Luo}, {Vito}, \& {Xue}}]{Ni2019}
{Ni}, Q., {Yang}, G., {Brandt}, W.~N., {et~al.} 2019, arXiv e-prints,
  arXiv:1909.06382

\bibitem[{{Nipoti} {et~al.}(2012){Nipoti}, {Treu}, {Leauthaud}, {Bundy},
  {Newman}, \& {Auger}}]{Nipoti2012}
{Nipoti}, C., {Treu}, T., {Leauthaud}, A., {et~al.} 2012, \mnras, 422, 1714

\bibitem[{{Peng} \& {Renzini}(2019)}]{Peng2019}
{Peng}, Y.-j., \& {Renzini}, A. 2019, arXiv e-prints, arXiv:1910.10446

\bibitem[{{Puglisi} {et~al.}(2019){Puglisi}, {Daddi}, {Liu}, {Bournaud},
  {Silverman}, {Circosta}, {Calabr{\`o}}, {Aravena}, {Cibinel}, {Dannerbauer},
  {Delvecchio}, {Elbaz}, {Gao}, {Gobat}, {Jin}, {Le Floc{\textquoteright}h},
  {Magdis}, {Mancini}, {Riechers}, {Rodighiero}, {Sargent}, {Valentino}, \&
  {Zanisi}}]{Puglisi2019}
{Puglisi}, A., {Daddi}, E., {Liu}, D., {et~al.} 2019, \apjl, 877, L23

\bibitem[{{Rangel} {et~al.}(2014){Rangel}, {Nandra}, {Barro}, {Brightman},
  {Hsu}, {Salvato}, {Koekemoer}, {Brusa}, {Laird}, {Trump}, {Croton}, {Koo},
  {Kocevski}, {Donley}, {Hathi}, {Peth}, {Faber}, {Mozena}, {Grogin},
  {Ferguson}, \& {Lai}}]{Rangel2014}
{Rangel}, C., {Nandra}, K., {Barro}, G., {et~al.} 2014, \mnras, 440, 3630

\bibitem[{{Rujopakarn} {et~al.}(2018){Rujopakarn}, {Nyland}, {Rieke}, {Barro},
  {Elbaz}, {Ivison}, {Jagannathan}, {Silverman}, {Smol{\v{c}}i{\'c}}, \&
  {Wang}}]{Rujopakarn2018}
{Rujopakarn}, W., {Nyland}, K., {Rieke}, G.~H., {et~al.} 2018, \apjl, 854, L4

\bibitem[{{Rujopakarn} {et~al.}(2019){Rujopakarn}, {Daddi}, {Rieke}, {Puglisi},
  {Schramm}, {P{\'e}rez-Gonz{\'a}lez}, {Magdis}, {Alberts}, {Bournaud},
  {Elbaz}, {Franco}, {Ivison}, {Kawinwanichakij}, {Kohno}, {Narayanan},
  {Silverman}, {Wang}, \& {Williams}}]{Rujopakarn2019}
{Rujopakarn}, W., {Daddi}, E., {Rieke}, G.~H., {et~al.} 2019, arXiv e-prints,
  arXiv:1904.04507

\bibitem[{{S{\'a}nchez} {et~al.}(2004){S{\'a}nchez}, {Jahnke}, {Wisotzki},
  {McIntosh}, {Bell}, {Barden}, {Beckwith}, {Borch}, {Caldwell},
  {H{\"a}ussler}, {Jogee}, {Meisenheimer}, {Peng}, {Rix}, {Somerville}, \&
  {Wolf}}]{Sanchez2004}
{S{\'a}nchez}, S.~F., {Jahnke}, K., {Wisotzki}, L., {et~al.} 2004, \apj, 614,
  586

\bibitem[{{Scholtz} {et~al.}(2018){Scholtz}, {Alexander}, {Harrison},
  {Rosario}, {McAlpine}, {Mullaney}, {Stanley}, {Simpson}, {Theuns}, {Bower},
  {Hickox}, {Santini}, \& {Swinbank}}]{Scholtz2018}
{Scholtz}, J., {Alexander}, D.~M., {Harrison}, C.~M., {et~al.} 2018, \mnras,
  475, 1288

\bibitem[{{Schulze} {et~al.}(2018){Schulze}, {Silverman}, {Kashino}, {Akiyama},
  {Schramm}, {Sanders}, {Kartaltepe}, {Daddi}, {Rodighiero}, \&
  {Renzini}}]{Schulze2018}
{Schulze}, A., {Silverman}, J.~D., {Kashino}, D., {et~al.} 2018, \apjs, 239, 22

\bibitem[{{Schulze} {et~al.}(2019){Schulze}, {Silverman}, {Daddi},
  {Rujopakarn}, {Liu}, {Schramm}, {Mainieri}, {Imanishi}, {Hirschmann}, \&
  {Jahnke}}]{Schulze2019}
{Schulze}, A., {Silverman}, J.~D., {Daddi}, E., {et~al.} 2019, arXiv e-prints,
  arXiv:1906.04290

\bibitem[{{Silverman} {et~al.}(2008){Silverman}, {Mainieri}, {Lehmer}, {Alexand
  er}, {Bauer}, {Bergeron}, {Brandt}, {Gilli}, {Hasinger}, {Schneider},
  {Tozzi}, {Vignali}, {Koekemoer}, {Miyaji}, {Popesso}, {Rosati}, \&
  {Szokoly}}]{Silverman2008}
{Silverman}, J.~D., {Mainieri}, V., {Lehmer}, B.~D., {et~al.} 2008, \apj, 675,
  1025

\bibitem[{{Tacchella} {et~al.}(2016){Tacchella}, {Dekel}, {Carollo},
  {Ceverino}, {DeGraf}, {Lapiner}, {Mand elker}, \& {Primack}}]{Tacchella2016}
{Tacchella}, S., {Dekel}, A., {Carollo}, C.~M., {et~al.} 2016, \mnras, 458, 242

\bibitem[{{Tacchella} {et~al.}(2018){Tacchella}, {Carollo}, {F{\"o}rster
  Schreiber}, {Renzini}, {Dekel}, {Genzel}, {Lang}, {Lilly}, {Mancini},
  {Onodera}, {Tacconi}, {Wuyts}, \& {Zamorani}}]{Tacchella2018}
{Tacchella}, S., {Carollo}, C.~M., {F{\"o}rster Schreiber}, N.~M., {et~al.}
  2018, \apj, 859, 56

\bibitem[{{van der Wel} {et~al.}(2012){van der Wel}, {Bell}, {H{\"a}ussler},
  {McGrath}, {Chang}, {Guo}, {McIntosh}, {Rix}, {Barden}, {Cheung}, {Faber},
  {Ferguson}, {Galametz}, {Grogin}, {Hartley}, {Kartaltepe}, {Kocevski},
  {Koekemoer}, {Lotz}, {Mozena}, {Peth}, \& {Peng}}]{vanderWel2012}
{van der Wel}, A., {Bell}, E.~F., {H{\"a}ussler}, B., {et~al.} 2012, \apjs,
  203, 24

\bibitem[{{van der Wel} {et~al.}(2014){van der Wel}, {Franx}, {van Dokkum},
  {Skelton}, {Momcheva}, {Whitaker}, {Brammer}, {Bell}, {Rix}, \&
  {Wuyts}}]{vanderWel2014}
{van der Wel}, A., {Franx}, M., {van Dokkum}, P.~G., {et~al.} 2014, \apj, 788,
  28

\bibitem[{{van Dokkum} {et~al.}(2008){van Dokkum}, {Franx}, {Kriek}, {Holden},
  {Illingworth}, {Magee}, {Bouwens}, {Marchesini}, {Quadri}, {Rudnick},
  {Taylor}, \& {Toft}}]{vanDokkum2008}
{van Dokkum}, P.~G., {Franx}, M., {Kriek}, M., {et~al.} 2008, \apjl, 677, L5

\end{thebibliography}





\end{document}